\documentclass{emulateapj}
\usepackage{amsmath}
\usepackage{graphicx}

\newcommand{\beq}{\begin{equation}}
\newcommand{\eeq}{\end{equation}}
\newcommand{\bet}{\vec{\beta}}
\newcommand{\nhat}{\hat{n}}
\newcommand{\dd}{\partial}

\begin{document}

\title{Pulsar Timing Constraints on Cumulative and Individual Mass of Stars in the Galactic Center}
\author{Pierre Christian and Abraham Loeb}
\affil{Harvard Smithsonian Center for Astrophysics}
\affil{60 Garden St, Cambridge, MA.}
\email{pchristian@cfa.harvard.edu}

\begin{abstract}
We consider the time derivatives of the period $P$ of pulsars at the Galactic Center due to variations in their orbital Doppler shifts. We show that in conjunction with a measurement of a pulsar's proper motion and its projected separation from the supermassive black hole, Sgr A*,measuring two of the three derivatives $\dot{P}$, $\ddot{P}$, or $\dddot{P}$ sets a constraint that allows for the recovery of the complete six phase space coordinates of the pulsar's orbit, as well as the enclosed mass within the orbit. Thus, one can use multiple pulsars at different distances from Sgr A* to determine the radial mass distribution of stars and stellar remnants at the Galactic center. Furthermore, we consider the effect of passing stars on the pulsar's period derivatives and show how it can be exploited to measure the characteristic stellar mass in the Galactic Center.
\end{abstract}

\section{Introduction}
The recent discovery of J1745-2900, a magnetar orbiting the supermassive black hole Sgr A* at a projected separation of $0.09$ pc \citep{b1,c3,c7} stimulated much interest in its timing and astrometry. Pulsars close to Sgr A* could allow for a precise measurement of the black hole's mass and spin, in addition to a host of relativistic effects \citep{c4,c10,c11,c5,c6}.

Unfortunately, the timing of the magnetar J1745-2900 is not sufficiently stable for dynamical measurements \citep{kaspi1}. Furthermore, it is located too far from Sgr A* (with a Keplerian orbital period of $\sim 500$ years) for it to be useful as a probe of strong field gravity. Calculations imply that there could be $\sim 200$ pulsars within a parsec from Sgr A* \citep{b6}, although perhaps only $\sim20$ of them being bright enough to be detected \citep{dexter}. Most of these pulsars might also be located too far from Sgr A* for testing strong field gravity. 

Nevertheless, one can still use pulsars at these larger distances to probe the astrophysical environment of the Galactic Center. In particular, the orbital dynamics of a pulsar is determined by the mass distribution within its orbit. Therefore, by measuring the imprint of the orbital Doppler effect on the pulsar's period, $P$, one should be able to constrain the radial mass profile of stars and stellar remnants around Sgr A*.  A previous study \citep{Gould} considered this possibility, but neglected the contributions of closely passing stars. In this letter, we evaluate the limitations of this technique due to this extra source of uncertainty, and also show that one can constrain the characteristic stellar mass in this environment by measuring the third time derivative of the pulsar's period, $\dddot{P}$.

This letter is organized as follows. In \S 2 we discuss the orbital contribution to the first, second, and third period derivatives $\dot{P}$, $\ddot{P}$, and $\dddot{P}$ by the mean field, and discuss how it can be used to measure the mass enclosed within the pulsar's orbit. In \S 3 we calculate the effects of passing stars on the period derivatives, and how it could be used to constrain the characteristic stellar mass in the Galactic Center.

\section{Measuring the enclosed mass via the mean field contribution to period derivatives}

\begin{figure}[t!]
\centering
\includegraphics[width=0.4\textwidth, trim=15 20 0 0,clip=true]{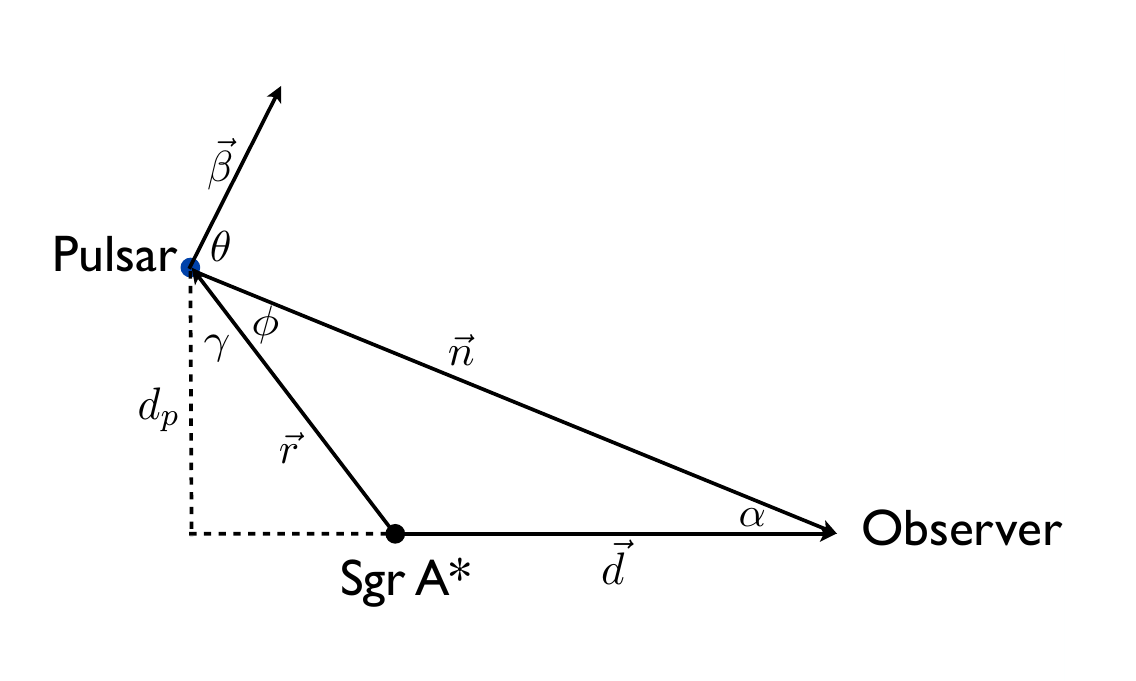}
\caption{The geometry under consideration. Note that the pulsar's velocity vector (in units of c), $\bet$, is not constrained to lie in the pulsar-Sgr A*-observer plane.}
\label{geometry}
\end{figure}

We begin with the equation for classical Doppler shift relating the observed period $P_O$ to the intrinsic (rest frame) period $P_i$:
\beq
P_O = P_i (1- \bet \cdot \nhat)  \; ,
\eeq
where $\bet$ is the velocity of the pulsar in units of $c$, and $\hat{n}$ is the unit vector pointing from the pulsar to the observer (see Figure \ref{geometry} for geometry). Since both $\vec{\beta}$ and $\hat{n}$ changes with time, the orbit of the pulsar induces nonzero time derivatives on $P_O$. Since the orbital time is much longer than the observation time, we can use a Taylor expansion in time $t$ to write,
\beq
P_O(t) = P_O(0) + \dot{P}_O(0) t + \frac{1}{2} \ddot{P}_O(0) t^2 + \frac{1}{6} \dddot{P}_O(0) t^3 + ... \; .
\eeq
The first two derivatives have been previously studied in the context of globular clusters \citep{phinney}. In this section we provide a treatment of $\dot{P}$, $\ddot{P}$, and $\dddot{P}$ for pulsars at the Galactic Center. In general, these time derivatives depend on the pulsar's 6 position and velocity phase space coordinates, as well as the enclosed mass. 

Direct imaging (e.g. \citealt{c3}) yields the projected separation of the pulsar from Sgr A*. The proper motions of pulsars have been measured previously both in the context of quantifying the pulsar's natal kick (e.g. \citealt{k2}) and for astrometric purposes (e.g. \citealt{k1}). In particular, the proper motion of J1745-2900 is currently being measured \citep{bower}.

Line of sight distances to pulsars can be obtained via parallax (see \citealt{k1} for an example of the technique applied to a millisecond pulsar \emph{not} at the Galactic Center), and progress has been made to measure the parallaxes of pulsars at large distances (up to $7.2$ kpc) using very-long-baseline interferometers \citep{b7}. The recently launched GAIA satellite\footnote{http://sci.esa.int/gaia/} is also expected to further improve the prospect of measuring pulsar parallaxes.

In addition to parallax, the distance to pulsars can be estimated from their pulse dispersion measure. The delay in pulse arrival time as a function of frequency, along with a model of the free electron distribution (e.g. the NE2001 model of \citealt{c1,c2}), can be used to estimate distances. This method was recently applied to the Galactic Center for the magnetar J1745-2900 \citep{c8,c9}. 

Based on the above measurements, one can determine 5 out of 6 of the pulsar's phase space coordinates. Another constraint can be placed via a measurement of one of the period derivatives, thereby constituting a full determination of its 6 phase space coordinates. Furthermore, we will show that with the measurement of another period derivative, one could directly measure the mass enclosed.

\subsection{The first period derivative}

The first time derivative of the pulsar's period $P$ is given by:
\beq \label{eq:tder1}
\dot{P}_O = -P_i \frac{\dd \bet \cdot \nhat }{\dd t} = - P_i \left( \bet \cdot \frac{\dd \nhat }{\dd t} + \nhat \cdot \frac{\dd \bet}{\dd t}  \right) \; ,
\eeq
where the subscript $O$ denotes the \emph{orbital} contribution to $\dot{P}$, in difference from the \emph{intrinsic} pulsar spindown, $\dot{P}_i$.
The acceleration of the pulsar is given by:
\beq
\frac{\dd \bet}{\dd t} = - \frac{G M }{r^2 c} \hat{r} \; ,
\eeq
where $M(r) \equiv M_{BH} + M_\star(r) $ is the total mass, namely the mass of the supermassive black hole Sgr A*, $M_{BH}=(4.31 \pm 0.36) \times 10^6 M_\odot$ \citep{massbh,ghez}, plus the mass of the stars within the pulsar's orbit. Substituting this into equation (\ref{eq:tder1}) gives:
\beq \label{eq:tder2}
\frac{\dot{P}_O}{P_i} = - \bet \cdot \frac{\dd \nhat}{\dd t} + \frac{G M }{r^2 c} (\nhat \cdot \hat{r}) = -\bet \cdot \frac{\dd \nhat}{\dd t} - \frac{G M}{r^2 c } \cos{\phi} \; ,
\eeq
where $\phi$ is the angle between the pulsar's radius vector $\vec{r}$ and the line of sight towards the pulsar, $\nhat$. The negative sign arises from the direction of $\hat{r}$. 
For $r/d \ll 1$, and taking account of the fact that $\vec{d}$ is time independent, 
\beq
\frac{\dd }{ \dd t} \nhat \approx -\frac{1}{d} \frac{\dd \vec{r}}{\dd t} = - \frac{c}{d} \bet \; ,
\eeq
where $\vec{d}$ is the displacement of the solar system barycenter from the supermassive black hole Sgr A*. 
Substituting this result in equation (\ref{eq:tder2}), yields
\beq
\frac{\dot{P}_O}{P_i} =\frac{c \beta^2 }{d} - \frac{G M }{r^2 c} \cos{\theta} = \frac{c \beta^2 }{d} - \frac{G M }{d_p^2 c} \sin^2{\phi} \cos{\phi} \; ,
\eeq 
where we have used the definition of the projected distance: $d_p \equiv r \cos{\gamma} \approx r \sin{\phi}$ in the second equality.
Solving for $\phi$, we obtain:
\beq
\frac{\sin^2{\phi}}{d_p^2} \cos{\phi} = \frac{c}{G M }\left [ \frac{c \beta^2}{d} - \frac{ \dot{P}_O}{P_i}  \right] \; .
\eeq
Defining $r_l$ as the component of $\vec{r}$ in the line of sight direction: 
\beq \label{eq:pdot}
\frac{r_l}{r^3} = \frac{c}{G M }\left [\frac{c \beta^2}{d} - \frac{ \dot{P}_O}{P_i} \right] \; .
\eeq
Note that in addition to the spindown rate due to the orbital Doppler effect, a portion of the observed $\dot{P}$ is due to intrinsic radiative losses. The overall spindown rate is the sum of the \emph{intrinsic} $\dot{P}_i$ and the orbital $\dot{P}_O$ components. Given this perspective, we view equation (\ref{eq:pdot}) as providing $\dot{P}_O$ as a function of $M(r)$ and the pulsar's phase space position: $\dot{P}_O(r,d_p,\beta,M)$. 
For a specific $\beta$, the value of $\dot{P}_O$ is bounded from above. For example, consider the magnetar J1745-2900 with an observed spindown rate of $\dot{P}_{obs}/P_i = 1.73 \times 10^{-12} \; \rm s^{-1}$ and $d_p = 0.09 \; \rm pc$ \citep{b1}. Approximating $M \approx M_{BH}$ and $c \beta \sim 150 \; \rm km \; s^{-1}$, we find that equation (\ref{eq:pdot}) obtains a maximum at  $\dot{P}_O/P_i \sim 10^{-13} \; \rm s^{-1}$. This means that the orbital contribution to $\dot{P}$ can account for at most $\sim 17\%$ of the observed $\dot{P}$ (assuming the observed $\dot{P}_{obs}$ value of \citealt{b1}). In this context, we can therefore be certain that measurements of the magnetar's magnetic field strength is not contaminated significantly by the orbital component $\dot{P}_O$.

\begin{figure}
\centering
\includegraphics[width=0.4\textwidth]{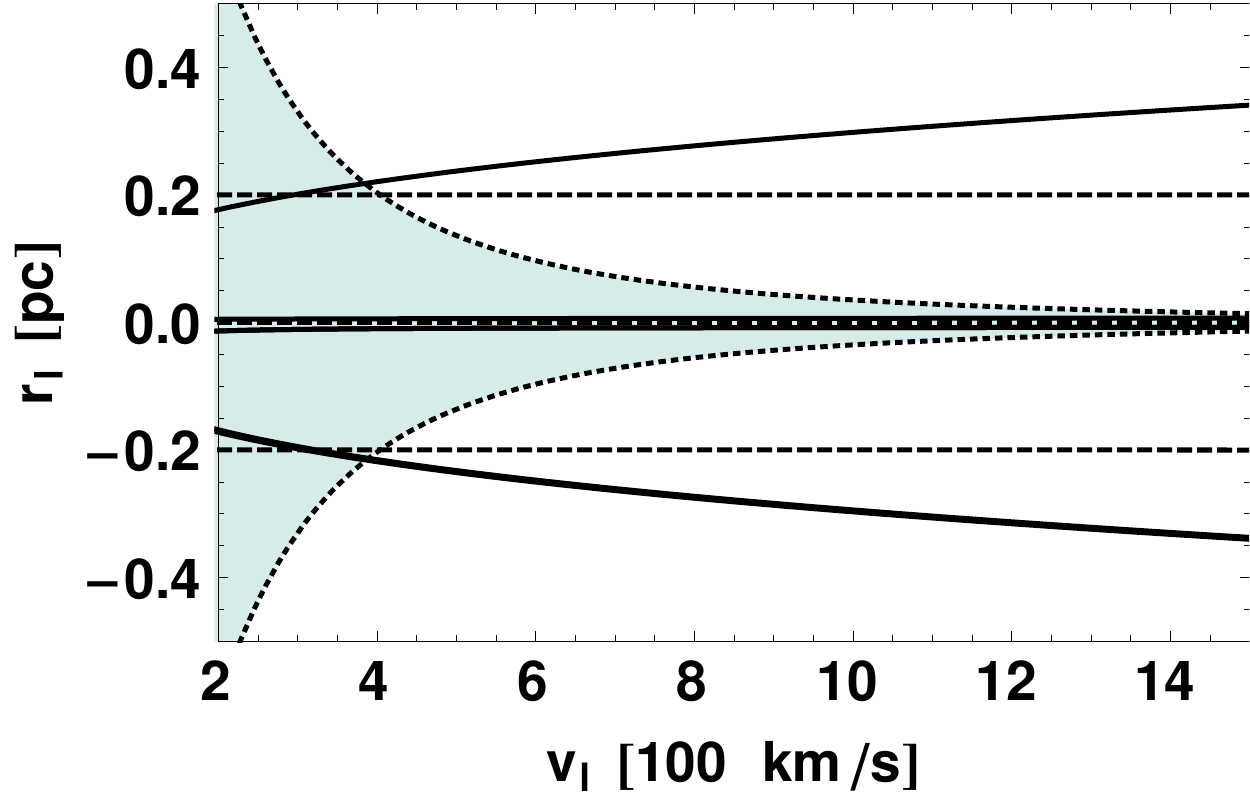}
\caption{Constraints on the line-of-sight componet of the pulsar's orbital radius $r_l$ and velocity $v_l$ based on $\dot{P}$ (dashed lines) and $\ddot{P}$ (solid lines)  for a case where the pulsar is orbiting in the Sgr A*-observer plane with $\dot{P} = 5 \times 10^{-15}$, $\ddot{P}=10^{-24} \; \rm s^{-1} $, $d_p = 0.01 \; \rm pc$, and $v_\perp = 150 \; \rm km \; s^{-1}$. Orbits in the shaded region are gravitationally bound to Sgr A*.}
\label{2constraint}
\end{figure}

If there is a way to measure $\dot{P}_O$ on its own (e.g. if the magnetic field of the pulsar is small, and the orbital contribution dominates), or in the case of millisecond pulsars, where $\dot{P}_O$ dominates $\dot{P}_{obs}$, equation (\ref{eq:pdot}) provides a new constraint to the pulsar's phase space position. For example, if it is observed that:
\beq
\frac{\dot{P}_O}{P_i} = \frac{c \beta^2}{d} \; ,
\eeq
then
\beq
r = d_p \; .
\eeq
Another case where the pulsar is orbiting in the Sgr A*-observer plane with  $\dot{P} = 5 \times 10^{-15}$ and $M\approx M_{BH}$ is presented in Figure \ref{2constraint}.

\subsection{The second period derivative}
If $\dot{P}_O$ is measured, equation (\ref{eq:pdot}) constitutes a new constraint on the pulsar's phase space coordinates, allowing all 6 components to be determined. This last constraint is a function of the enclosed mass, $M$, which can be solved via a measurement of another period derivative. Taking the derivative of equation (\ref{eq:tder1}), we find:
\begin{align} \label{eq:Pddotsuperraw}
\frac{\ddot{P}_O}{P_i} &= - \bet \cdot \frac{\partial^2 \hat{n}}{\partial t^2} - \frac{\partial \bet}{\partial t} \cdot \frac{\partial \hat{n}}{\partial t} -\frac{G M}{c} \frac{\dd}{\dd t} \left[ \frac{1}{r^2} \cos\phi    \right]  \; .
\end{align}
Noting that
\begin{align}
-\bet \cdot \frac{\dd^2 \hat{n}}{\dd t^2}=  \bet \cdot \frac{ c}{d } \frac{\dd \bet}{\dd t} = - \frac{G M}{r^2 d} \bet \cdot \hat{r} \; ,
\end{align}
and
\beq
- \frac{\partial \bet}{\partial t} \cdot \frac{\partial \hat{n}}{\partial t} = \frac{G M}{r^2 d} \hat{r} \cdot \bet \; ,
\eeq
the first and second terms of the right-hand-side of equation (\ref{eq:Pddotsuperraw}) cancel, leaving:
\begin{align}
\frac{\ddot{P}_O}{P_i} &= - \frac{G M}{c}\frac{\dd }{\dd t}\left[ \frac{\cos\phi}{r^2} \right]
\\ &= \frac{GM}{c}\left[ \frac{\sin\phi}{r^2}\frac{\dd \phi}{\dd t} + \frac{2}{r^3}\cos \phi \frac{\dd r}{\dd t} \right] \; . \label{eq:raw}
\end{align}
A pulsar on a purely circular orbit with an orbital frequency $\Omega=\beta c/r$ orbiting in the pulsar-Sgr A*-observer plane, has
\beq
\frac{\dd r}{\dd t} = 0 \;\;\;\;\;\;\;\; ; \;\;\;\;\;\;\;\;  \frac{\dd \phi}{\dd t} = -\Omega = -\beta \frac{c}{r} \; ,
\eeq
implying
\beq
\frac{\ddot{P}_O}{P_i} = -\frac{G M}{c} \frac{\sin\phi}{r^2} \Omega = -G M   \frac{d_p  }{ r^4} \beta \; ,
\eeq
which can be solved trivially for either $M(r)$ or $r(M)$ given $\beta$ (or limits of the quantity given the proper motion, $\beta_\perp \le \beta$) and the projected separation $d_p$. For a pulsar at the projected separation of J1745-2900 ($d_p=0.09$ pc), that is currently at a phase of its orbit where $r \sim d_p$ and $\beta \sim 0.3 \times 10^{-3}$, we find that:
\beq
|\ddot{P}| \approx 5 \times 10^{-23} \rm s^{-1} \; .
\eeq
Within 1 year, the drift in $\dot{P}$ is:
\beq
|\Delta \dot{P}| \approx 1.5 \times 10^{-15} \; ,
\eeq
which is within the precision attainable in pulsar measurements \citep{b2,b8}. If a millisecond pulsar is found at the Galactic Center, then $\dot{P}$ could be measured to a precision of $10^{-20}$ (e.g. \citealt{b9}). In general, $r^2 = (d_p^2 + r_l^2)$, therefore:
\begin{align}
\frac{\dd r}{\dd t} &= \frac{1}{2r}\frac{\dd r^2}{\dd t}
\\&= \frac{1}{\pm 2 \sqrt{d_p^2 + r_l^2} } \frac{\dd}{\dd t} (d_p^2 + r_l^2) 
\\&= \pm \frac{1}{r} \left(d_p v_\perp \frac{\vec{v}_\perp \cdot \vec{d}_p}{d_p v_\perp} + r_l v_l\right) \; ,
\end{align}
where $\vec{v}_\perp$ is the proper velocity and $v_l$ is the velocity in the line of sight direction. Furthermore we note that:
\begin{align}
-\frac{\sin\phi}{r^2} \frac{\dd \phi}{\dd t} &= \frac{d_p}{r^4} v \cos i \; ,
\end{align}
where $i$ is the inclination of the pulsar's orbit. This factor of $\cos i$ can be written as:
\beq
\cos i = \pm \frac{\sqrt{v^2 - v_{oop}^2}}   {v} = \pm \frac{ \sqrt{v_\perp^2 + v_l^2 - v_{oop}^2}}   {v} \; ,
\eeq
where,
\beq
v_{oop} = \frac{|\vec{v}_\perp \times \vec{d}_{p}| }{d_p}
\eeq
is the component of the proper velocity $v_\perp$ that is off the pulsar-Sgr A*-observer plane. This quantity can be obtained directly from the proper velocity by a projection to the pulsar-Sgr A*-observer plane. Combining, we obtain,
\begin{align} \label{eq:result}
\frac{\ddot{P}_O}{P_i} &= \frac{G M }{c(d_p^2 + r_l^2)^2 } \left[ \pm 2 \frac{r_l}{\sqrt{d_p^2 + r_l^2}} \left(\vec{v}_\perp \cdot \vec{d}_p + r_l v_l\right) \right. \nonumber \\ 
&\quad \quad \left. \pm d_p\sqrt{v_\perp^2 + v_l^2 - v_{oop}^2} \right] \; .
\end{align}
Measuring $\ddot{P}_O/P_i$ provides another constraint on the orbital phase space coordinates and the enclosed mass of the pulsar's orbit. An example of constraining $r_l(v_l)$ where the pulsar is orbiting in the pulsar-Sgr A*-observer plane with $\ddot{P}=10^{-24}\; \rm s^{-1}$ and $M\approx M_{BH}$ is plotted in Figure \ref{2constraint}. 

\begin{figure}
\centering
\includegraphics[width=0.4\textwidth]{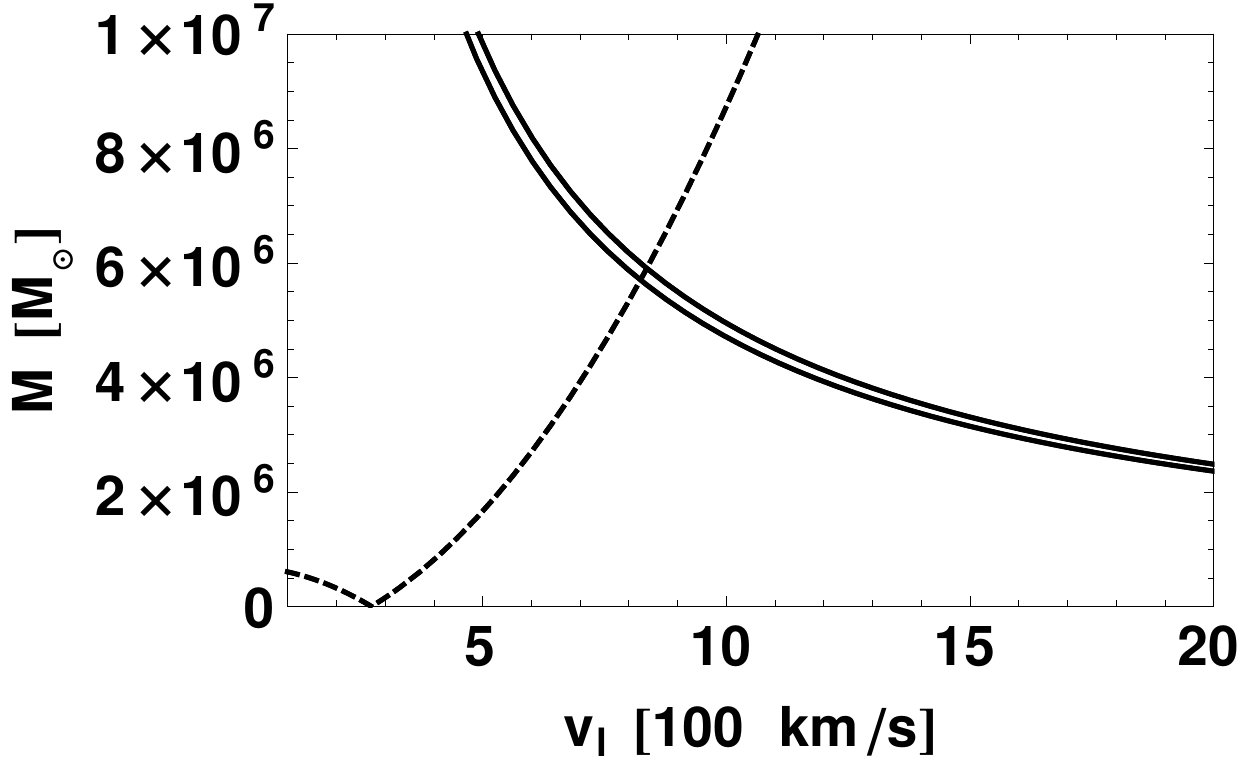}
\caption{Constraints on the mass distribution based on $\ddot{P}_O$ and $\dot{P}_O$. Supposing that the other 5 phase space coordinates are known, the constraints from $\ddot{P}_O$ (solid line) and $\dot{P}_O$ (dashed line) can be solved simultaneously for $v_l$ and $M$, corresponding to the intersection of the two constraints. In this case, the pulsar is orbiting in the pulsar-Sgr A*-observer plane with $\dot{P} = 10^{-14}$, $\ddot{P}=10^{-23} \; \rm s^{-1} $, $d_p = 0.01 \; \rm pc$, $v_\perp = 150 \; \rm km \; s^{-1}$, and $r_l = 0.2 \; \rm pc$. Due to the sign degeneracy, the $\ddot{P}$ constraint in this case corresponds to two solid lines in the $M$ versus $v_l$ plane. This degeneracy can be resolved by fixing $v_l$, for example, through the third derivative constraint. Note that enclosed masses below the mass of Sgr A*, $M_{BH}=(4.31 \pm 0.36) \times 10^6 M_\odot$ \citep{ghez} should be excluded from the analysis.}
\label{mass}
\end{figure}

If we instead assume that $r_l$ is known (e.g. from parallax), the constraints from $\ddot{P}_O$ and $\dot{P}_O$ can be used to solve simultaneously for $v_l$ and the mass enclosed within the orbital radius. This corresponds to the intersection of the two constraints in Figure \ref{mass}. 

\subsubsection{The intrinsic $\ddot{P}$}
While the observed $\ddot{P}$ of a young pulsar is typically dominated by its intrinsic timing noise, the intrinsic $\ddot{P}$ for old or millisecond pulsars can be dominated instead by radiation losses \citep{liv}. Assuming the vacuum dipole model, we can estimate the intrinsic contribution, $\ddot{P}_i$. If we assume that $B \sin\alpha$ does not change significantly with time (i.e. that magnetic field decay timescales are long), the quantity $(P\dot{P})_{i}$ is constant:
\beq
(P \dot{P})_{i} = \frac{8 \pi^2 R^6 (B \sin \alpha)^2 }{3c^3 I} = \rm constant \; ,
\eeq
where $\alpha$ is the magnetic axis inclination angle, $B$ the magnetic field strength, $I$ the moment of inertia, and $R$ the neutron star radius. Taking another time derivative, we obtain:
\beq
\ddot{P}_{i} = -\frac{64 \pi^4 R^{12} (B \sin\alpha)^4}{9 c^9 I^2 P_i} \; .
\eeq
Considering $P \approx P_i \approx 1 \; \rm s$, a magnetic field $B= 10^{12} \; \rm G$, and the typical radius, $R = 10 \; \rm km$, and mass, $M=1.4 M_{\odot}$, for a neutron star, and adopting $\sin\alpha=1$, we obtain the \textbf{maximum} $\ddot{P}_{i}$ to be:
\beq
\ddot{P}_{i, \; \rm max} = 7.6 \times 10^{-31} \; \rm s^{-1} \; ,
\eeq
which is very small compared to the orbital contribution. This suggests that unlike $\dot{P}$, $\ddot{P}$ is much less contaminated by the intrinsic contribution, allowing a clean measurement of $\ddot{P}_O$.
In general, the inclination angle $\alpha$ can be inferred by various methods \citep{b2,b3,b4,b5}. 

\subsection{The Third period Derivative}
Since the most well studied pulsars have measured $\dddot{P}$ \citep{b8}, we now supply the third period derivative. This third timing constrain can be used in lieu of the $\dot{P}$ constraint in cases where $\dot{P}_{obs}$ is too contaminated by $\dot{P}_i$. In addition, all three timing constraints can be used in cases where only 4 phase space coordinates are measured (e.g. for pulsars with no measured parallax). We also note that the third derivative $\dddot{P}$ is even less affected by its intrinsic contribution than $\ddot{P}$, since $\dddot{P}_i \propto B^6/P_i^3$.  The derivative of equation (\ref{eq:raw}) is:

\begin{align} \label{eq:raw3}
\frac{\dddot{P}_O}{P_i} &=   \frac{G M }{c} \left[ \frac{2}{r^3} \cos \phi \frac{\dd^2 r}{\dd t^2} + \frac{1}{r^2} \sin \phi \frac{\dd^2 \phi}{\dd t^2} \right. \nonumber \\
&\left. -\frac{6}{r^4} \cos\phi \left(\frac{\dd r}{\dd t} \right)^2  +\frac{1}{r^2} \cos \phi \left(\frac{\dd \phi}{\dd t} \right)^2 -\frac{4}{r^3} \sin \phi \frac{\dd r}{\dd t}\frac{\dd \phi}{\dd t} \right] \; ,
\end{align}

with the following equalities: 
\begin{align}
\frac{\dd^2 r}{\dd t^2} &= -\frac{2}{r^2}(d_p v_\perp + r_l v_l) \frac{\dd r}{\dd t} + \frac{1}{r} \left[ v_\perp \frac{\vec{d}_p \cdot \vec{v}_\perp  }{d_p} \right. \nonumber \\
&\left. \quad - d_p \frac{a \sin \phi}{v_\perp}\frac{\vec{d}_p \cdot \vec{v}_\perp  }{d_p} + v_l^2 + r_l a \cos \phi   \right] \; ,
\end{align}
\begin{equation}
\frac{\dd^2 \phi}{\dd t^2} = \frac{\dd}{\dd t} \left(- \frac{v}{r} \cos i \right) 
 = \cos i \left[\frac{v}{r^2} \frac{\dd r}{\dd t} - \frac{1}{r} \frac{\dd v}{\dd t} \right] \;,
\end{equation}
and 
\beq
\frac{\dd v}{\dd t} = \frac{1}{v} \left[ v_l a \cos \phi - v_\perp \frac{a \sin \phi}{v_\perp} \frac{\vec{d}_p \cdot \vec{v}_\perp }{d_p} \right] \; ,
\eeq
where $a = GM/r^2$. For a well studied pulsar with both a measured $\ddot{P}$ and $\dddot{P}$, we get two timing constraints on the phase space coordinates. If the pulsar has a low magnetic field strength or if $\dot{P}_O$ dominates, there will be a third timing constraint. Treating $M(r)$ as an unknown function, these extra constraints can be used to directly measure the mass enclosed within the pulsar's orbit. The method is analogous to that presented in \S 2.3.

\section{Contributions by stellar kicks and measuring the characteristic stellar mass at the Galactic Center}
As pointed out in \cite{phinney}, stars passing close to the pulsar can gravitationally kick the pulsar, adding another contribution to the time derivatives of the pulsar's period. In this section we quantify the probability for these stochastic effects to significantly affect the mean field contribution. The probability of a star being a distance $< b$ away from the pulsar located a distance $r$ away from Sgr A* is:
\beq \label{eq:prob1}
Pr(r) = 1 - \exp[-(4 \pi /3) n_* \pi b^3] \; ,
\eeq
where $n_*$ is the number density of stars. Using a density profile for stars around Sgr A* $n_* = n_0 (r/r_0)^{-1.8}$, and the fact that the total stellar mass at $1 \; \rm pc$ is measured to be $M_{1pc}\sim 2 \times 10^6 \; M_\odot$ \citep{genzel}, the probability represented in equation (\ref{eq:prob1}) becomes:
\beq \label{eq:prob2}
Pr(r) = 1 - \exp \left[- \frac{4 \pi}{3} b^3 \left( \frac{1.2 M_{1pc}}{4 \pi m_* (1 \; \rm pc )^{1.2} }   \right) r^{-1.8} \right]  \; .
\eeq
The contribution of the nearest neighboring star to $\dot{P}$ equals the mean field contribution at a distance that satisfies,
\beq
b^2 = \frac{m_*}{M_*(r) + M_{BH}} r^2 \; ,
\eeq
where $M_*(r)$ is the total stellar mass within the orbit. The probability for this separation is,
\begin{align}
&Pr(r)= 1 \nonumber \\
&\quad - \exp\left[ - \frac{1.2}{3} \frac{M_{1pc} \sqrt{m_*}}{(M_{1pc}(r/1 \; \rm pc)^{1.2} +M_{BH})^{3/2} } \left(\frac{r}{1 \; \rm pc} \right)^{1.2}  \right] \; .
\end{align}
For all reasonable values for $m_*$, this probability is negligibly small at all radii, showing that the first derivative is uncontaminated by stellar kicks. We corroborated this analytical analysis with a numerical N-body simulation utilizing the Salpeter mass function for stars. The initial conditions for this simulation were generated using the star cluster integrator, \emph{bhint} \citep{bhint} with density profile $\rho \propto r^{-1.8}$ and a supermassive black hole of mass $4.3 \times 10^6 M_\odot$ located at the center of the cluster.

\begin{figure}[t!]
\centering
\includegraphics[width=0.4\textwidth]{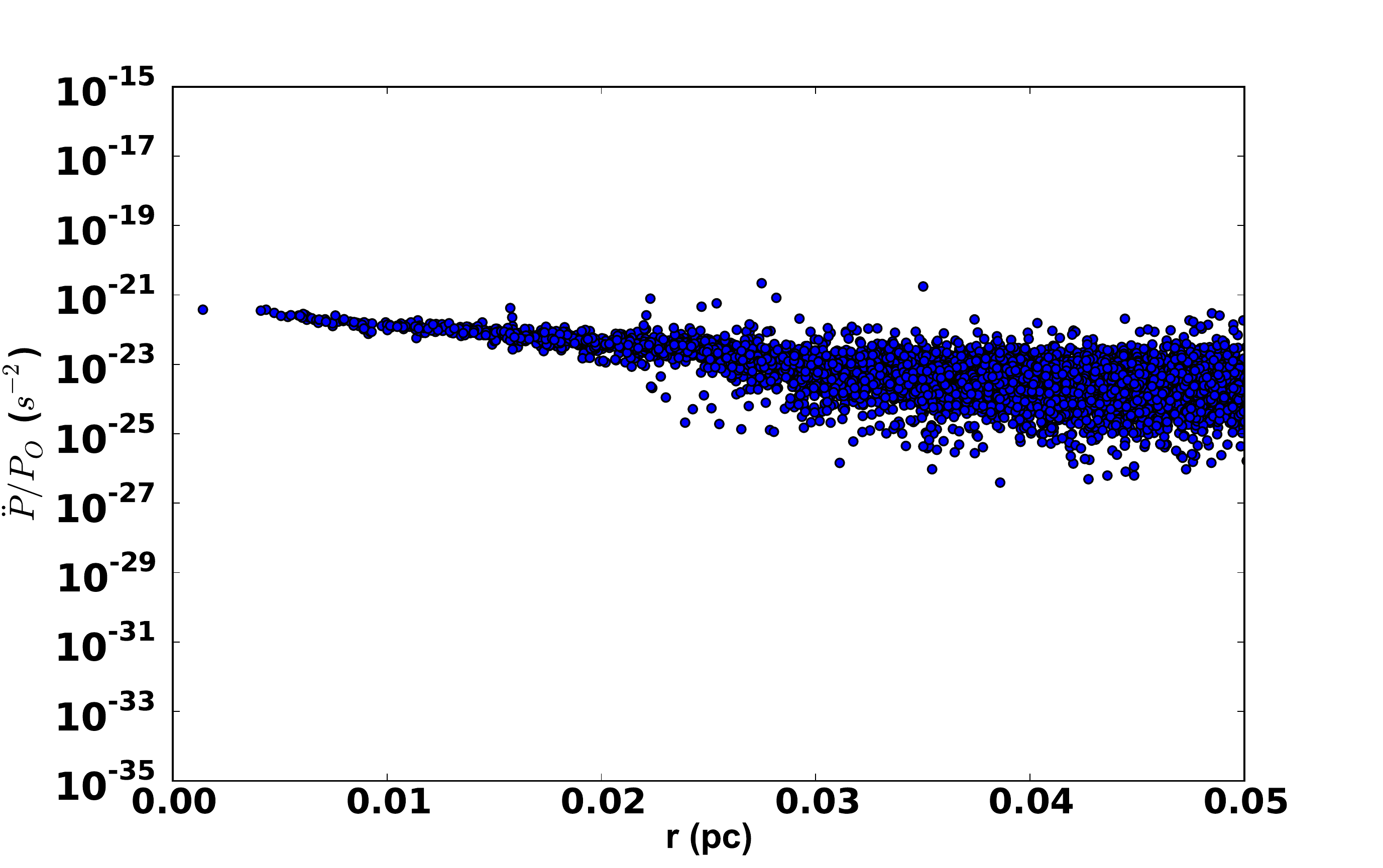}
\caption{Numerical simulation of $\ddot{P}/P_O$. The Galactic Center was modeled as a star cluster generated by the post-Newtonian integrator \emph{bhint} \citep{bhint}   with density profile $\rho \propto r^{-1.8}$ and a  central supermassive black hole of mass $4.3 \times 10^6 M_\odot$.} 
\label{numerical}
\end{figure}

However, the contributions of stellar kicks to the higher derivatives is larger. The nearest neighbor contribution to $\ddot{P}/P$ is \citep{phinney}:
\beq
\left[\frac{\ddot{P} }{P}\right]_{nn} \approx \frac{G m_* }{b^3} \frac{v_*}{c} \; ,
\eeq
where $v_*$ is the relative velocity between the star and the pulsar. Similarly, the mean field contribution is:
\beq
\left[\frac{\ddot{P} }{P}\right]_{mf} \approx \frac{G (M_*(r)+M_{BH})}{r^3} \frac{v}{c} \; , 
\eeq
where $v$ is the pulsar's orbital speed relative to Sgr A*. Equating the two contributions, we find that the nearest neighbor contribution equals the mean field at,
\beq
b^3 = \frac{m_*}{M_*(r)+M_{BH}} \frac{v_*}{v} r^3 \; .
\eeq
The probability for this separation is again obtained from equation (\ref{eq:prob2}):
\begin{align}
&Pr(r) = 1 \nonumber \\
&\quad -\exp \left[ - \frac{1.2}{3} \frac{v_*}{v} \frac{M_{1pc}}{M_{1pc}  (r/1 \; \rm pc)^{1.2} + M_{BH} } \left(\frac{r}{1 \; \rm pc} \right)^{1.2} \right] \; .
\end{align}
The probability decreases with $r$, so that pulsars closer to Sgr A* are less disturbed by perturbing stars. At a distance of $0.01$ pc, the probability for the associated jerks is less than $0.1\%$. To corroborate this result, we performed a numerical simulation utilizing the Salpeter mass function as displayed in Figure \ref{numerical}. This simulation shows that passing stars scatter $\ddot{P}$ about its mean field value, and its contribution is large at distances larger than $\sim 0.03 $ pc. 

While the analysis of \S 2 is largely unaffected at small $r$, the contribution of passing stars adds a significant source of uncertainty to the interpretation of measured period derivatives at large $r$. Due to the stochastic nature of this contribution, measuring $M(r)$ at large distances necessitates the use of multiple pulsars, whose average $\ddot{P}$ should reveal the mean field.

\subsection{Effect of changing $m_*$ on $\dddot{P}$}

\begin{figure}[h!] 
\centering
\includegraphics[width=0.4\textwidth]{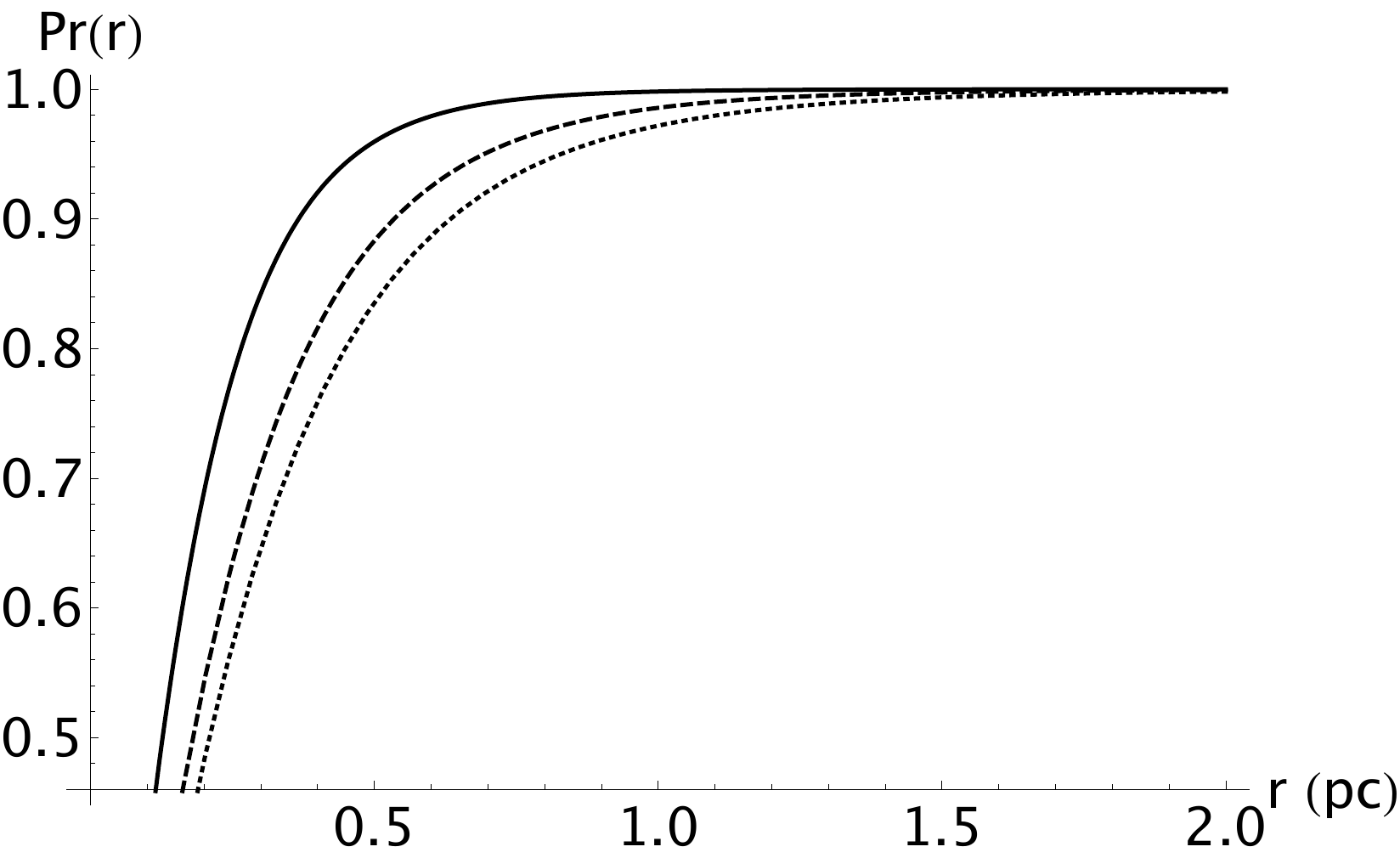}
\caption{\label{thirdder} The probability for a large nearest neighbor contribution to $\dddot{P}/P$ versus pulsar's orbital radius $r$ for characteristic stellar masses of $m_*=1, \;5,\; \& \; 10 M_\odot$ in solid, dashed, and dotted lines, respectively.}
\end{figure}

Although the probability for $\ddot{P}$ to be significantly contaminated by perturbing stars is independent of the characteristic stellar mass, $m_*$, this is not true for all time derivatives. Here we show that the third period derivative is sensitive to changes in $m_*$, and that this dependence can be used to probe the characteristic stellar mass at the Galactic Center. The nearest neighbor contribution to $\dddot{P}/P$ is \citep{phinney}:
\begin{align}
\left[\frac{\dddot{P} }{P}\right]_{nn} \approx 2 \frac{G m_*}{b^4} \frac{v_*^2}{c} \; .
\end{align}
The mean field contribution is:
\beq
\left[\frac{\dddot{P} }{P}\right]_{mf} \approx 2 \left[ \frac{G (M_* (r)+M_{BH}) }{r^4} \right] \frac{v^2}{c} \; .
\eeq
The nearest neighbor and the mean field contributions are equal when,
\beq
b^3 =  \left[ \frac{m_*}{M_*(r)+M_{BH}} \left(\frac{v_*}{v}\right)^2 r^4 \right]^{3/4} \; .
\eeq
In this case, the probability for a star to pass at distance $<b$ from the pulsar is:
\begin{align}
Pr(r) &= 1- \exp \left[  -  \frac{1.2}{3 m_* ^{1/4} } \left( \frac{v_*}{v} \right)^{3/2}  \right. \nonumber \\
&\quad \left. \times \frac{M_{1pc}}{(M_{1pc} (r/1 \; \rm pc)^{1.2} + M_{BH} )^{3/4} } \left(\frac{r}{1 \; \rm pc} \right)^{1.2}  \right] \; .
\end{align}
Figure \ref{thirdder} shows this probability as a function of pulsar-Galactic Center distance for $m_*=1, \;5,\; \& \; 10 M_\odot$. The differences between cases of different $m_*$'s maxes out at around $r\approx 2 $ pc, thus making this the optimal location to perform this study. We note that the difference between $Pr(0.2 \; \rm pc)$ with $m_*=1M_\odot$ and $m_*=10M_\odot$ is $\sim20\%$, and so with Poisson statistic one needs tens of pulsars to distinguish between the two cases.

\section{Summary and Implications}
If the pulsar's projected separation from Sgr A*, proper velocity, and line of sight distance are measured, then 5 of the 6 pulsar's phase space coordinates are known. One of the period derivative constraints represented by equations (\ref{eq:pdot}), (\ref{eq:result}), or (\ref{eq:raw3}) can then be used to  derive a final constraint on the pulsar's phase space coordinates. Another constraint can be used to limit the mass enclosed within the pulsar's orbit. In this case, the line of sight velocity $v_l=v_l(M)$ depends on $M$ itself. As such, $M(\vec{r}, \vec{v})=M(\vec{r}, \vec{v}_\perp, v_l(M,\vec{r}, \vec{v}_\perp))=M(\vec{r}, \vec{v}_\perp)$. Due to this complicated dependence, an analytic solution is not feasible and the related analysis has to be done numerically.

The mass distribution itself can be determined if the above measurements are performed at multiple times for a pulsar on a plunging orbit. However, unless the pulsar is located very close to Sgr A*, the orbital timescale is too long for such a study. Nevertheless, by performing this measurement on multiple pulsars, one would still be able to probe the radial mass distribution. In particular, the difference in the measured $M(r)$'s of two pulsars located at two radial distances determines the mass enclosed in the spherical shell between these radii. This can be used to constrain the distribution of low-mass stars or stellar remnants (black holes, white dwarfs, and neutron stars) that are too faint to be detected directly.

An extra source of uncertainty in measuring $M(r)$ comes from the effects of passing stars. These scatter $\ddot{P}$ about the mean field value, and the contribution is large for $r$ greater than $\sim 0.03$ pc. As such, measurements should be taken close to Sgr A*, where they can constrain cumulative mass of stars and stellar remnants surrounding the black hole. Measuring $M(r)$ further away from Sgr A* will require multiple pulsars, and thus be a challenging task. 

Finally, we note that the scatter of $\dddot{P}$ about the mean field value due to passing stars is affected by the characteristic stelar mass, $m_*$. As such, measurements of $\dddot{P}$ of multiple pulsars at the Galactic Center will allow us to probe the charasteristic mass of stars and remnants in this extreme environment. Such measurements can also place exquisite constraints on the existence of intermediate-mass black holes in the vicinity of Sgr A*.

\section{Acknowledgment}
We thank Vicky Kaspi and Sterl Phinney for helpful comments on the manuscript. This work was supported in part by NSF grant AST-1312034.


\begin{thebibliography}{99}
\bibitem[\protect\citeauthoryear{Bower et al.}{2014}]{bower} Bower, G. C. et al. 2014,
ApJ, 780, L2
\bibitem[\protect\citeauthoryear{Champion et al.}{2005}]{b9} Champion, D. J. et al. 2005,
MNRAS, 363, 929
\bibitem[\protect\citeauthoryear{Chaname and Gould}{2002}]{Gould} Chaname, J.; Gould, A. 2002,
ApJ, 571, 320
\bibitem[\protect\citeauthoryear{Chatterjee et al.}{2009}]{b7}  Chatterjee, S. et al. 2009,
ApJ, 698, 250
\bibitem[\protect\citeauthoryear{Chennamangalam and Lorimer}{2013}]{b6}  Chennamangalam, J.; Lorimer, D. R. 2013,
MNRAS, tmpL, 44C
\bibitem[\protect\citeauthoryear{Cordes et al.}{2002}]{c10} Cordes, J. M. et al. 2002,
New AR, 48, 1413
\bibitem[\protect\citeauthoryear{Cordes and Lazio}{2002}]{c1} Cordes, J. M.; Lazio, T. J. W. 2002,
astro-ph/0207156
\bibitem[\protect\citeauthoryear{Cordes and Lazio}{2003}]{c2} Cordes, J. M.; Lazio, T. J. W. 2003,
astro-ph/0301598
\bibitem[\protect\citeauthoryear{Dexter and O'Leary}{2014}]{dexter} Dexter, J.; O'Leary, R. 2014,
ApJ, 782, L38
\bibitem[\protect\citeauthoryear{Du et al.}{2014}]{k1} Du, Yuanjie et al. 2014,
ApJ, 782, L38
\bibitem[\protect\citeauthoryear{Eatough et al.}{2013}]{c8} Eatough, R. P. et al. 2013,
Nature, 501, 391
\bibitem[\protect\citeauthoryear{Genzel et al.}{2010}]{genzel} Genzel, R. et al. 2010,
Reviews of Modern Physics, 82, 3121
\bibitem[\protect\citeauthoryear{Ghez et al.}{2008}]{ghez} Ghez, A. M. et al. 2008,
ApJ, 689, 1044
\bibitem[\protect\citeauthoryear{Gillessen et al.}{2009}]{massbh} Gillessen, S. et al. 2009,
ApJ, 692, 1075
\bibitem[\protect\citeauthoryear{Kaplan et al.}{2008}]{k2} Kaplan, D. L. et al. 2008,
ApJ, 677, 1201
\bibitem[\protect\citeauthoryear{Kaspi et al.}{2014}]{kaspi1} Kaspi, V. M. et al. 2014,
astro-ph: 1403.5344
\bibitem[\protect\citeauthoryear{Kennea et al.}{2013}]{c7} Kennea, J. A. et al. 2013,
ApJ, 770, L24
\bibitem[\protect\citeauthoryear{Kramer et al.}{2004}]{c11} Kramer, M. et al. 2004,
New AR, 48, 993
\bibitem[\protect\citeauthoryear{Liu et al.}{2012}]{c6} Liu, K. et al. 2012,
ApJ, 747, 1
\bibitem[\protect\citeauthoryear{Lorimer}{2008}]{liv} Lorimer, D. R.  2008,
Living Rev. Relativity 11, 8, accessed April 2014
\bibitem[\protect\citeauthoryear{Lockmann and Baumgardt}{2008}]{bhint}  Lockmann, U.; Baumgardt, H. 2008,
MNRAS, 384, 323
\bibitem[\protect\citeauthoryear{Lyne and Manchester}{1993}]{b3}  Lyne, A. G.; Manchester, R. N. 1988,
MNRAS, 234, 477, LM88
\bibitem[\protect\citeauthoryear{Manchester et al.}{2005}]{b8} Manchester, R. N. et al. 2005,
AJ, 129, 1993
\bibitem[\protect\citeauthoryear{Miller and Hamilton}{1993}]{b5}  Miller, M. C.;Hamilton, R. J. 1993,
ApJ, 411, 298, 301
\bibitem[\protect\citeauthoryear{Mori et al.}{2013}]{b1} Mori K. et al. 2013,
ApJ, 770, L23
\bibitem[\protect\citeauthoryear{Pfahl and Loeb}{2004}]{c4} Pfahl E., Loeb A. 2004,
ApJ, 615, 253
\bibitem[\protect\citeauthoryear{Phinney}{1993}]{phinney} Phinney, E. S. 1993,
ASPC, 50, 141
\bibitem[\protect\citeauthoryear{Psaltis and Johannsen}{2010}]{c5} Psaltis D., Johannsen T. 2010,
JPhCS, 283, 2030P
\bibitem[\protect\citeauthoryear{Rankin}{1990}]{b4}  Rankin, J. M. 1990,
ApJ, 352, 247, R90
\bibitem[\protect\citeauthoryear{Rea}{2013}]{c3} Rea, N. et al. 2013,
ApJl, 775, L34
\bibitem[\protect\citeauthoryear{Shannon and Johnston}{2013}]{c9} Shannon R. M., Johnston S., 2013,
MNRAS, 435, L29
\bibitem[\protect\citeauthoryear{Taylor, Manchester, and Lyne}{1993}]{b2} Taylor, J. H.; Manchester, R. N.; Lyne, A. G. 1993,
ApJ Supplement Series, 88, 2, 529-568
\end{thebibliography}
\end{document}